\documentclass[10pt,technote]{article}
\usepackage{amsmath}
\usepackage{amsthm}
\usepackage{latexsym}
\usepackage{amssymb}
\usepackage{xspace}
\usepackage{amscd}
\usepackage{enumerate}

\newcommand{\PAut}{{\rm PAut}}

\newcommand{\Gal}{{\rm Gal}}

\newcommand{\Aut}{\mbox{\rm Aut}}

\newcommand{\Imagen}{\mbox{\rm Im}}

\newcommand{\inv}{^{-1}}
\newcommand{\F}{\mathbb F}

\newcommand{\K}{{\mathbb K}}
\newcommand{\I}{{\mathcal I}}
\newcommand{\G}{\mathcal{G}}

\newcommand{\GL}{{\rm GL}}

\newcommand{\matriz}[1]{\begin{array} #1 \end{array}}

\newcommand{\GEN}[1]{\langle #1 \rangle}

\title{Group code structures on affine-invariant codes}

\author{Jos\'{e} Joaqu\'{\i}n Bernal, \'{A}ngel del R\'{\i}o and Juan Jacobo Sim\'{o}n
\footnote{\hspace{-0.4cm} Departamento de Matem\'{a}ticas, Universidad de Murcia, 30100
Murcia. Spain. \newline email: josejoaquin.bernal@alu.um.es, adelrio@um.es,
jsimon@um.es\newline Partially supported by D.G.I. of Spain and Fundaci\'{o}n S\'{e}neca of
Murcia.}}

\newtheorem{theorem}{Theorem}
\newtheorem{lemma}[theorem]{Lemma}
\newtheorem{proposition}[theorem]{Proposition}

\newtheorem{definition}[theorem]{Definition}
\newtheorem{corollary}[theorem]{Corollary}
\newtheorem{remark}[theorem]{Remark}

\newtheorem{notation}[theorem]{Notation}
\newtheorem{example}[theorem]{Example}

\theoremstyle{remark}
\theoremstyle{remark}

\begin{document}

\maketitle

\begin{abstract}
A group code structure of a linear code is a description of the code as one-sided or two-sided ideal of a group algebra
of a finite group. In these realizations, the group algebra is identified with the ambient space, and the group
elements with the coordinates of the ambient space. It is well known that every affine-invariant code of length $p^m$,
with $p$ prime, can be realized as an ideal of the group algebra $\F\I$, where $\I$ is the underlying additive group of
the field with $p^m$ elements. In this paper we describe all the group code structures of an affine-invariant code of
length $p^m$ in terms of a family of maps from $\I$ to the group of automorphisms of $\I$.
\end{abstract}

Affine-invariant codes were firstly introduced by Kasami, Lin and Peterson \cite{KLP2} as a generalization of
Reed-Muller codes. This class of codes has received the attention of several authors because of its good algebraic and
decoding properties \cite{D,BCh,ChL,Ho,Hu}. The length of an affine-invariant code is a prime power $p^m$, where $p$ is
the characteristic of the finite field $\F$ which plays the role of alphabet. It is well known that every
affine-invariant code of length $p^m$ over the field $\F$ can be realized as an ideal of the group algebra $\F\I$,
where $\I$ is the underlying additive group of the field with $p^m$ elements, i.e. $\I$ is the elementary abelian group
of order $p^m$. In this realization, the group elements are identified with the elements of the standard base of the
ambient space $\F^{p^m}$. We refer to these realizations of codes as one-sided or two-sided ideals in group algebras as
group code structures of the given code. In this paper we study all the possible group code structures on an
affine-invariant code.

Our main tools are an intrinsical characterization of group codes obtained in \cite{BRS} and a description of the group
of permutation automorphisms of non-trivial affine-invariant codes given in \cite{BCh}. These results are reviewed in
Section 1, where we also recall the definition and main properties of affine-invariant codes. In Section~\ref{Carac},
we describe all the group code structures of an affine-invariant code $C$ in terms of a family of maps $\I\rightarrow
\G_{a,b}$ where $\G_{a,b}$ is a subgroup of the group of automorphism of $\I$ depending on two integers $a$ and $b$
which are determined by the code $C$. Some methods to calculate $a$ and $b$ were given in \cite{D} and \cite{BCh}. As
an application we exhibit in Section~\ref{Ejemplos} a family of group code structures on any affine-invariant code $C$
for which the integer $a$ is different from $m$ and characterize the affine-invariant codes $C$ which have a
non-abelian group code structure.

\section{Preliminaries}

In this section we recall the definition of (left) group code and the intrinsical characterization given in \cite{BRS}.
We also recall the definition of affine-invariant code and the description of its group of permutation automorphisms
given in \cite{BCh}.

All throughout $p$ is a positive prime integer and all the fields in this paper are
finite of characteristic $p$. The field with $p^s$ elements is denoted by
$\F_{p^s}$. Two finite fields $\F$ and $\K=\F_{p^m}$ (of characteristic $p$) are
going to be fixed throughout the paper. The roles of alphabet and length of the
affine-invariant codes of this paper are going to be represented by $\F$ and $p^m$,
respectively. We denote the underlying additive group of $\K$ with $\I$. We will
abuse of the notation and some maps defined on $\I$ (respectively, $\K$) will be
considered acting in $\K$ (respectively, $\I$) too. We use this double notation to
emphasize whether we are considering the additive or the field structure.

The group algebra of a group $G$ with coefficients in the field $\F$ is going to be denoted by $\F G$. In particular
$\F\I$ denotes the group algebra of the underlying additive group of $\K$ with coefficients in $\F$ and $\F(\K^*)$ is
the group algebra of the group of units of $\K$ with coefficients in $\F$. To avoid the ambiguity of an expression of
the form $\sum_{g\in \I} a_g g$, as either an element of $\F\I$ or an element of a field containing both $\F$ and $\K$,
we use the notation $\sum_{g\in \I} a_g \overline{g}$ to represent elements of $\F\I$. We consider the group algebra
$\F(\K^*)$ as a subspace of $\F \I$ (not as a subalgebra). All the group theoretical notions used in this paper can be
easily founded in \cite{R}.

\begin{definition}
If $E=\{e_1,\dots,e_n\}$ is the standard basis of $\F^n$, $C\subseteq \F^n$ is a linear code and
$G$ is a group (of order $n$) then we say that $C$ is a {\em left $G$-code} (respectively, a {\em
right $G$-code}; a {\em $G$-code}) if there is a bijection $\phi:\{1,2,\dots,n\}\rightarrow G$ such
that the linear map $\overline{\phi}:\F^n\rightarrow \F G$, given by $e_i\mapsto \phi(i)$, maps $C$
to a left ideal (respectively, a right ideal; a two-sided ideal) of $\F G$.

A {\em left group code} (respectively, {\em group code}) is a linear code which is a left $G$-code (respectively, a $G$-code) for some group $G$.

A (left) cyclic group code (respectively, abelian group code, solvable group code, etc.) is
a linear code which is (left) $G$-code for some cyclic group $G$ (respectively, abelian
group, solvable group, etc.).
\end{definition}

Let $S_n$ denote the group of permutations on $n$ symbols. Every $\sigma\in S_n$ defines an
automorphism of $\F^n$ in the obvious way, i.e.
$\sigma(x_1,\dots,x_n)=(x_{\sigma\inv(1)},\dots,x_{\sigma\inv(n)})$. By definition, the group of
permutation automorphisms of a linear code $C$ of length $n$ is
    \begin{equation}\label{PAut1}
    \PAut(C)=\{\sigma\in S_n : \sigma(C)=C\}.
    \end{equation}

An intrinsical characterization of (left) group codes $C$ in terms of $\PAut(C)$ has been obtained in \cite{BRS}.

\begin{theorem}\label{characterization} \cite{BRS}
Let $C$ be a linear code of length $n$ over a field $\F$ and let $G$ be a finite group of order
$n$.
\begin{enumerate}[(a)]
\item $C$ is a left $G$-code if and only if $G$ is isomorphic to a transitive subgroup of $S_n$ contained in
$\PAut(C)$.
\item
$C$ is a $G$-code if and only if $G$ is isomorphic to a transitive subgroup $H$ of $S_n$
such that $H\cup C_{S_n}(H)\subseteq \PAut(C)$, where $C_{S_n}(H)$ denotes the centralizer
of $H$ in $S_n$.
\end{enumerate}
\end{theorem}

Recall that $\I$ denotes the underlying additive group of $\K=\F_{p^m}$. Let
$S(\I)$ denote the group of bijections $\I\rightarrow \I$. Every element of $S(\I)$
induces a unique $\F$-linear bijection of the group algebra $\F\I$.
Affine-invariant codes of length $p^m$ are defined as subspaces of the group
algebra $\F\I$, rather than subspaces of $\F^{p^m}$, in terms of its group of
permutation automorphisms considered as elements of $S(\I)$. We explain now the
transfer from $\PAut$ as a subgroup of $S_{p^m}$ to the group of permutation
automorphisms as elements of $S(\I)$.

For an $\F$-subspace $C$ of $\F \I$, let
    \begin{equation}\label{PAut2}
    \PAut(C)=\{\sigma \in S(\I):\sigma(C)=C\}.
    \end{equation}
Observe that if $\phi:\{1,\dots,p^m\}\rightarrow \I$ is a bijection and $D$ is a linear code in
$\F^{p^m}$ then $\phi$ induces an isomorphism $\PAut(D)\simeq \PAut(\overline{\phi}(D))$, where the
left side $\PAut$ uses (\ref{PAut1}) and the second one uses (\ref{PAut2}). Therefore, if $C$ is a
subspace of $\F\I$ and $G$ is a group of order $p^m$ then $C$ is a left $G$-code if and only if
$\PAut(C)$ contains a transitive subgroup $H$ of $S(\I)$ isomorphic to $G$ and it is a $G$-code if
$H$ can be selected such that $C_{S(\I)}(H)\subseteq \PAut(C)$.

\begin{definition}
An affine-invariant code is an $\F$-subspace $C$ of $\F\I$ such that $\PAut(C)$ contains the maps of the form $x\in \I
\mapsto \alpha x+\beta$, with $\alpha \in \K^*$ and $\beta\in \I$ and every element $\sum_{g\in \I} a_g\overline{g}$ of
$C$ satisfies $\sum_{g\in \I} a_g=0$.
\end{definition}

Affine-invariant codes can be seen as extended cyclic codes as follows. If $J$ is an ideal of
$\F(\K^*)$ then the parity check extension of $J$ is
    $$\left\{\sum_{g\in \I} a_g \overline{g} : \sum_{g\in \K^*} a_g \overline{g} \in J \text{ and }
    \sum_{g\in \K^*} a_g=0\right\}.$$
If $C\subseteq \F\I$ is an affine-invariant code then $C^* = \{\sum_{g\in \K^*} a_g \overline{g} :
\sum_{g\in \I} a_g \overline{g} \in C\}$ is an ideal of $\F(\K^*)$ and $C$ is the parity check
extension of $C^*$.

We recall a characterization of Kasami, Lin and Peterson of the parity check
extensions of ideals of $\F(\K^*)$ which are affine-invariant in terms of the
$p$-adic expansion of its defining set \cite{KLP1}. Let $C\subseteq \F\I$ be the
parity check extension of an ideal of $\F(\K^*)$. The defining set of $C$ is
    $$D_C = \left\{ i: \matriz{{l} 0\le i < p^m \text{ and} \\
    \sum_{g\in \I} a_g g^i = 0, \text{for every } \sum_{g \in \I} a_g \overline{g} \in C}\right\},$$
where, by convention, $0^0=1$. If $q$ is the cardinality of $\F$ then $D_C\setminus\{p^m-1\}$ is a union of
$q$-cyclotomic classes modulo $p^m-1$. Conversely, if $D$ is a subset of $\{0,1,\dots,p^m-1\}$, such that $D\setminus
\{p^m-1\}$ is a union of $q$-cyclotomic classes modulo $p^m-1$, then there is a unique ideal $J$ of $\F(\K^*)$ such
that $D$ is the defining set of the parity check extension of $J$ (see e.g. \cite{Ch}).

The $p$-adic expansion of a non-negative integer $x$ is the list of integers $(x_0,x_1,\dots)$,
uniquely defined by $0\le x_i < p$ and $x=\sum_{i\ge 0} x_i p^i$. The $p$-adic expansion yields a
partial ordering in the set of positive integers by setting $x\preceq y$ if $x_i\le y_i$, for every
$i$, where $(x_i)$ and $(y_i)$ are the $p$-adic expansions of $x$ and $y$, respectively.

\begin{proposition}\cite{KLP1}\cite[Corollary~3.5]{Hu}
Let $C\subseteq \F\I$ be the parity check extension of an ideal $\F(\K^*)$. Then $C$ is
affine-invariant if and only if $D_C$ satisfies the following condition for every $1\le s,t\le
p^m-1$:
    \begin{equation}\label{Del}
    s\preceq t \text{ and } t\in D_C \quad \Rightarrow \quad s\in D_C.
    \end{equation}
\end{proposition}

Three obvious affine-invariant codes are the zero code, the repetition code and its dual, i.e. $\{0\}$, the ideal of
$\F\I$ generated by $\sum_{g\in \I} \overline{g}$, and the augmentation ideal $\{\sum_{g\in \I} a_g \overline{g} :
\sum_{g\in \I} a_g=0\}$, respectively. Their defining sets are $\{0,1,\dots,p^m-1\}$, $\{0,1,\dots,p^m-2\}$ and
$\{0\}$, respectively. These three codes are known as the trivial affine-invariant codes \cite{BCh,Hu}.

For future use we describe the affine-invariant codes of length 4.

\begin{example}[Affine-invariant codes of length 4]\label{Ex4}

{\rm Let $D$ be the defining set of an affine-invariant code of length $4$ over
$\F_{2^r}$. Thus $D$ satisfies condition (\ref{Del}) and $D\setminus \{3\}$ is a
union of $2^r$ classes modulo $3$. If $r$ is even then the $2^r$-cyclotomic classes
modulo 3 are $\{0\}$ and $\{1,2\}$. This implies that if $r$ is even then
$D=\{0\}$, $\{0,1,2\}$ or $\{0,1,2,3\}$, i.e. $C$ is trivial as affine-invariant
code. However, if $r$ is odd then the cyclotomic classes modulo $3$ are $\{0\},
\{1\}$ and $\{2\}$. So, in this case there are two additional possibilities for
$D$, namely $\{0,1\}$ and $\{0,2\}$. Resuming, if $r$ is even then there are not
$2^r$-adic non-trivial affine-invariant codes of length $4$ and if $r$ is odd then
there are two $2^r$-adic non-trivial affine invariant codes of length $4$.
 }\end{example}

If $C$ is a trivial affine-invariant code then $\PAut(C)=S_n$, and therefore $C$ is $G$-code for every group $G$ of
order $p^m$. So to avoid trivialities, in the remainder of the paper all the affine-invariant codes are suppose to be
non-trivial. The group of permutations of a (non-trivial) affine-invariant code has been described by Berger and
Charpin \cite{BCh}.

We identify every element $y\in \I$ with the translation $x\mapsto x+y$, so that the group $\I$ can be identified with
the group of translations of $\K$. If $L/E$ is a field extension then $\Gal(L/E)$ denotes the Galois group of $L$ over
$E$, i.e. the group of field automorphisms of $L$ which fix the elements of $E$. To refer to $L$ as a vector space over
$E$ we write $L_E$. Accordingly $\GL(L_E)$ denotes the groups of linear transformations of $L$ as vector space over
$E$.

\begin{theorem}\label{PAut}\cite[Corollary~2]{BCh}
Let $C$ be a non-trivial affine-invariant code of length $p^m$ over $\F=\F_{p^r}$ and let $K=\F_{p^m}$. Let
    $$\matriz{{l}
    a=a(C) = \min \left\{ d|m : \GL(\K_{\F_{p^d}})\subseteq \PAut(C) \right\} \text{ and} \\
    b=b(C) = \min \left\{ d\ge 1 : \matriz{{l} D_C\setminus\{p^m-1\} \text{ is a union} \\
        \text{of  cyclotomic } p^d-\text{classes} \\\text{modulo } p^m-1}\right\}.
    }$$
Then $b|r$, $b|a|m$ and
    $$\PAut(C)=\GEN{\I,\GL(\K_{\F_{p^a}}),\Gal(\K/\F_{p^b})}.$$
\end{theorem}

A method to compute $a(C)$ and $b(C)$ was firstly obtained by Delsarte \cite{D}. Later, Berger and Charpin gave two
alternative methods which are sometimes computationally simpler \cite{BCh}.

Now we present an alternative description of
$\PAut(C)=\GEN{\I,\GL(\K_{\F_{p^a}}),\Gal(\K/\F_{p^b})}$, for $C$ an
affine-invariant code as in Theorem~\ref{PAut}. We use the notation $N\rtimes G$ to
represent a semidirect product of $N$ by $G$ via some action of $G$ on $N$. That
is, $N$ and $G$ are two groups and there is a group homomorphism
$\sigma:G\rightarrow \Aut(N)$. The map $\sigma$ is referred to as the action of the
semidirect product and in most examples it will be clear from the context. We use
$\sigma_g=\sigma(g)$, to minimize the number of parenthesis. The underlying set of
$N\rtimes G$ is the direct product $N\times G$ and the product is given by
$(n_1,g_1)(n_2,g_2)=(n_1\sigma_{g_1}(n_2),g_1g_2)$.

Given two divisors $a$ and $b$ of $m$ with $b|a$, let
    $$\G_{a,b} = \left\{f\in \GL(\K_{\F_{p^b}}) : \matriz{{l} f \text{ is } \tau-
    \text{semilinear for some } \\ \tau \in \Gal\left(\F_{p^a}/\F_{p^b}\right)}\right\}.$$
We claim that $\G_{a,b}=\GEN{\GL(\K_{\F_{p^a}}),\Gal(\K/\F_{p^b})}$. Indeed, if $f$ is $\tau$-semilinear with $\tau\in
\Gal(\F_{p^a}/\F_{p^b})$ then $\tau$ is the restriction of $\sigma$ for some $\sigma\in \Gal(\K/\F_{p^b})$. Then
$f\sigma\inv \in \GL(\K_{\F_{p^a}})$. This proves one inclusion; the other one is obvious. Using that $\I$ (identified
with the group of translations of $\K$), is normalized by $\GL(K_{\F_p})$ we deduce the following from
Theorem~\ref{PAut}.

\begin{corollary}\label{PAutSemi}
If $C$ is an affine-invariant code as in Theorem~\ref{PAut} then $\PAut(C)=\I\rtimes \G_{a,b}$.
\end{corollary}

\begin{remark}\label{mapT}{\rm
The map $T:\G_{a,b}\rightarrow \Gal(\F_{p^a}/\F_{p^b})$ which associates $f$ to $\tau$, when $f$ is $\tau$-semilinear,
is a surjective group homomorphism with kernel $\GL(\K_{\F_{p^a}})$.}
\end{remark}

\section{Group code structures on affine-invariant codes}\label{Carac}

In this section we present the main result of the paper, namely a description of all the group code structures of a
non-trivial affine-invariant code $C$ with $a=a(C)$ and $b=b(C)$ in terms of some maps $\alpha:\I\rightarrow \G_{a,b}$.

Given a map $\alpha:\I\rightarrow \G_{a,b}$ let
    $$\I_{\alpha} = \{(x,\alpha(x)\inv): x\in \I\}.$$ The proof of the following lemma is straightforward.

\begin{lemma}\label{IalphaGrupo}
$\I_{\alpha}$ is a subgroup of $\I\rtimes \G_{a,b}$ if and only if
    \begin{equation}\label{IYB}
    \alpha(x+y)=\alpha(\alpha(y)(x))\alpha(y)
    \end{equation}
for every $x,y\in \I$. In particular, if $\alpha$ satisfies (\ref{IYB}) then $\{\alpha(x):x\in \I\}$ is a $p$-subgroup
of $\G_{a,b}$.
\end{lemma}

We need one more lemma which is an easy consequence of Sylow's Theorem \cite[1.6.16]{R}.

\begin{lemma}\label{suma imagenes}
If $P$ is a $p$-subgroup of $\GL(\K_{\F_{p^a}})$ then $\sum_{\rho\in P} \Imagen(\rho-1)\ne \K$.
\end{lemma}
\begin{proof}
Select a basis $b_1,\dots,b_n$ of $\K_{\F_{p^a}}$, with $n=m/a$, and let $U$ be the set of endomorphisms $f$ of
$\K_{\F_{p^a}}$ such that for every $i=1,2,\dots,n$, $f(b_i)-b_i$ belongs to the $\F_{p^a}$-subspace of $\K$ generated
by the $b_j$'s with $1\leq j<i$. That is, $U$ is the group of automorphisms of $\K_{\F_{p^a}}$ having upper
unitriangular associated matrix in the given basis. An easy counting argument shows that $|U|=p^{a\frac{n(n-1)}{2}}$
and $|\GL(\K_{\F_{p^a}})|=(p^a-1)(p^{na}-p^a)(p^{na}-p^{2a})\dots
(p^{na}-p^{(n-1)a})=|U|(p^a-1)(p^{(n-1)a}-1)(p^{(n-2)a}-1)\dots (p^{a}-1)$. Then $U$ is a Sylow's $p$-subgroup of
$\GL(\K_{\F_{p^a}})$. By Sylow Theorem, there is $g\in \GL(\K_{\F_{p^a}})$ such that $P\subseteq gUg\inv$. Then
$\sum_{\rho\in P} \Imagen(\rho -1) \subseteq \sum_{u\in U} \Imagen(gug\inv-1) \subseteq g(\sum_{u\in u}
\Imagen(u-1))=g(\GEN{b_1,\dots,b_{n-1}})\ne \K$.
\end{proof}

\bigskip We are ready to present our main result.

\begin{theorem}\label{affinegroupcodes}
Let $\F$ and $\K$ be finite fields of characteristic $p$ and let $\I$ be the underlying additive group of $\K$. Let
$C\subseteq \F\I$ be a non-trivial affine-invariant code and let $a=a(C)$ and $b=b(C)$. Then the following assertions
hold for every finite group $G$:
\begin{enumerate}[(a)]
\item $C$ is a left $G$-code if and only if $G$ is isomorphic to $\I_{\alpha}$ for some map
$\alpha:\I\rightarrow \G_{a,b}$ satisfying condition (\ref{IYB}).
\item $C$ is a $G$-code if and only if $G$ is isomorphic to $\I_{\alpha}$ for some map
$\alpha:\I\rightarrow \GL(\K_{\F_{p^a}})$ satisfying condition (\ref{IYB}) and such that the map $\beta:\K\times \K
\rightarrow \K$ given by $\beta(x,y)=\alpha(x)\inv(y)-y$ is $\F_{p^a}$-bilinear.
\end{enumerate}
\end{theorem}

\begin{proof}
(a) First of all note that for every $0\neq y\in \I$ there exist
$(y,\alpha(y)^{-1})\in \I_{\alpha}\subseteq S(\I)$ such that
$(y,\alpha(y)^{-1})(0)=y+\alpha(y)^{-1}(0)=y$. Hence, if $\alpha$ satisfies
condition (\ref{IYB}) then $\I_{\alpha}$ is a transitive subgroup of $S(\I)$. Since
$|\I_{\alpha}|=|\I|=p^m$, the sufficiency follows from
Theorem~\ref{characterization}.

Conversely, assume that $C$ is a left $G$-code for some group $G$, necessarily of
order $p^m$. By Theorem~\ref{characterization} and Corollary \ref{PAutSemi}, we may
assume without loss of generality that $G$ is a transitive subgroup of $S(\I)$
contained in $\I\rtimes \G_{a,b}$. Thus, if $(x,g)$ and $(y,h)$ are two different
elements of $G$ with $x,y\in \I$ and $g,h\in \G_{a,b}$ then $x=(x,g)(0)\ne
(y,h)(0)=y$, and this shows that the projection of $G$ onto $\I$ is bijective. If
$\lambda:\I\rightarrow G$ is the inverse of this bijection then
$G=\{(x,\lambda(x))|x\in \I\}$. Define $\alpha(x):=\lambda(x)\inv$, for any $x\in
\I$. Then $G=\I_{\alpha}$ and $\alpha$ satisfies condition (\ref{IYB}), by
Lemma~\ref{IalphaGrupo}.

(b) Let $\alpha:\I \rightarrow \GL(\K_{\F_{p^a}})$ satisfy the conditions of (b).
By part (a) and Theorem~\ref{characterization} and Corollary \ref{PAutSemi} to
prove that $C$ is an $\I_{\alpha}$-code it is enough to show that
$C_{S(I)}(\I_{\alpha})\subseteq \I\rtimes \G_{a,b}$. In fact we are going to show
that $C_{S(I)}(\I_{\alpha})\subseteq\I\rtimes \GL(\K_{\F_{p^a}})$.

For every $x\in \I$ we set $\lambda(x)=\alpha(x)\inv$. Since $\I_{\alpha}$ is a
transitive subgroup of $S(\I)$ of order $|\I|=p^m$, the centralizer of $\I_{\alpha}$
in $S(\I)$ is $C_{S(\I)}(\I_{\alpha})=\{f(x):x\in \I\}$, where
    $$f(x)(y)=(y,\lambda(y))(x)=y+\lambda(y)(x).$$
(See \cite[Lemma~1.1]{BRS}, specialized to $i_0=0$.) For every $x,y\in \I$ set
    $$\lambda'(x)(y)=y+\lambda(y)(x)-x=y+\beta(y,x).$$
We claim that $\lambda'(x)\in \GL(\K_{\F_{p^a}})$, for every $x\in \I$. Indeed, on the one hand
$\lambda'(x)$ is the composition of $f(x)$ and the translation $y\mapsto y-x$. This shows that
$\lambda'(x)$ is bijective. On the other hand, since $\beta$ is $\F_{p^a}$-linear, if $y,y_1,y_2\in
\I$ and $\gamma\in \F_{p^a}$ then $\lambda'(x)(y_1+y_2) = y_1+y_2+\beta(y_1+y_2,x) =
y_1+\beta(y_1,x)+y_2+\beta(y_2,x) =\lambda'(x)(y_1)+\lambda'(x)(y_2)$ and  $\lambda'(x)(\gamma y) =
\gamma y+\beta(x,\gamma y) = \gamma (y + \beta(x,y)) =  \gamma \lambda'(x)(y)$. So
$C_{S(\I)}(\I_{\alpha})=\{(x,\lambda'(x)):x\in \I\}\subseteq \I\rtimes \GL(\K_{\F_{p^a}})$ as
wanted.

Conversely, assume that $C$ is a $G$-code. By Theorem~\ref{characterization} and the first part, $G\simeq \I_{\alpha}$
for a map $\alpha:\I\rightarrow \G_{a,b}$ satisfying condition (\ref{IYB}) and such that $C_{S(\I)}(\I_{\alpha})
\subseteq \PAut(C) = \I\rtimes \G_{a,b}$. We set $\lambda(x)=\alpha(x)\inv$ and $\beta(x,y)=\lambda(x)(y)-y$. We have
to show that $\beta$ is $\F_{p^a}$-bilinear. As in the previous paragraph the centralizer of $\I_{\alpha}$ in $S(\I)$
is formed by the maps $f(x):y\in \I \mapsto y+\lambda(y)(x)$, with $x\in \I$. So $f(x)\in \I\rtimes \G_{a,b}$, for
every $x\in \I$. Since $f(x)(0)=x$, we have $f(x)=(x,\lambda'(x))$, with $\lambda'(x)(y)=y+\lambda(y)(x)-x$. In other
words, $C_{S(\I)}\left(\I_\alpha\right)=\I_{\alpha'}$ for $\alpha'(x)=\lambda'(x)\inv$. By Lemma~\ref{IalphaGrupo},
$\alpha'$ satisfies condition (\ref{IYB}). Then $\I_{\alpha}$ and $\I_{\alpha'}$ are the centralizer of each other in
$S(\I)$ and their roles and the roles of $\lambda$ and $\lambda'$ can be interchanged. Since
$\beta(x,y)=\lambda(x)(y)-y=\lambda'(y)(x)-x$, to prove that $\beta$ is $\F_{p^a}$-bilinear it is enough to show that
$\lambda(x),\lambda'(x)\in \GL(\K_{\F_{p^a}})$ for every $x$. By symmetry, we only prove that $\lambda'(x)\in
\GL(\K_{\F_{p^a}})$, for every $x\in \I$.

Using that $\lambda'(y)$ is additive for every $y\in \I$, we have the following equality for every $x_1,x_2\in \I$:
    $$\matriz{{l}
    x_1+x_2+\lambda(x_1+x_2)(y)-y=\lambda'(y)(x_1+x_2)\\
     = \lambda'(y)(x_1)+\lambda'(y)(x_2) \\=
    x_1+\lambda(x_1)(y)-y+x_2+\lambda(x_2)(y)-y.}$$
Thus
    \begin{equation}\label{Lambda+}
    \lambda(x_1+x_2)+1=\lambda(x_1)+\lambda(x_2).
    \end{equation}

Let $Q$ be the subgroup generated by the $\lambda(x)$'s. Since $Q$ is a $p$-subgroup of $\G_{a,b}$,
by Remark \ref{mapT}, $T(Q)$ is a $p$-subgroup of $\Gal(\F_{p^a}/\F_{p^b})$. Let $P=Q\cap
\GL(\K_{\F_{p^a}})$. We fix a transversal $\mathcal{T}$ of $P$ in $Q$ containing $1$, and for every
$x\in \I$ we put $\delta(x)=\lambda(x)-t_x$, where $t_x$ is the only element of $\mathcal{T}$ with
$\lambda(x){t_x}\inv \in P$. Define $J=\sum_{x\in \I} \Imagen(\delta(x))$. Then
    \[J= \sum_{x\in \I}\Imagen\left[\left(\lambda(x)t^{-1}_x-1\right)t_x\right]
    \subseteq \sum_{\rho\in P} \Imagen(\rho-1) \ne \K\]
by Lemma~\ref{suma imagenes}.

For every $x\in \I$, let $\tau_x=T(\lambda(x))$ and $\tau'_x=T(\lambda'(x))$, i.e. $\lambda(x)$ is
$\tau_x$-semilinear and $\lambda'(x)$ is $\tau'_x$-semilinear. Observe that the condition
$\lambda(x)t_x\inv\in P$ is equivalent to $\lambda(x){t_x}\inv\in \GL(\K_{\F_{p^a}})$ and hence
$t_x$ is $\tau_x$-semilinear. Having in mind that $\lambda'(x)(y)=y+\lambda(y)(x)-x$ and
$\lambda(x)=\delta(x)+t_x$ we have
    \[\lambda'(x)(\gamma y)=\gamma y+\delta(\gamma y)(x)+t_{\gamma y}(x)-x\]
and
    \[\matriz{{rcl}
    \lambda'(x)(\gamma y)&=&\tau'_x(\gamma)\lambda'(x)(y) \\ &=&
    \tau'_x(\gamma)\left(y+\delta (y)(x)+t_y(x)-x\right)}\]
for any $\gamma \in \F_{p^a}$, and $x,y \in \I$. Therefore
    \begin{equation}\label{Igualdad}
    \matriz{{rcl} (\gamma -\tau'_x(\gamma)) y &=& x-t_{\gamma y}(x)  +\tau'_x(\gamma)(t_y(x)-x) + \\ &&
    \delta(y)({\tau_y}\inv\tau'_x(\gamma) x)- \delta(\gamma y)(x).}
    \end{equation}

Recall that the goal is proving that $\lambda'(x)\in \GL(\K_{\F_{p^a}})$, or
equivalently that $\tau'_x\ne 1$, for every $x\in \I$. By means of contradiction
assume that this is not the case and fix an element $x$ in $\I$ such that
$\tau'_x\ne 1$. We also fix $\gamma\in \F_{p^a}$ with $\tau'_x(\gamma)\ne \gamma$.
Observe that the order of $\tau'_x$ is a power of $p$ because so is the order of
$G$. Therefore $p$ divides $a/b$.

For every $y \in \I$ let
    $$z_y = x-t_{\gamma y}(x)  +\tau'_x(\gamma)(t_y(x)-x)$$
and set
    $$Z=Z_{x,\gamma}=\{z_y:y\in \I\}.$$
 Using (\ref{Igualdad}), we have
    $$\I=\{(\gamma -\tau'_x(\gamma)) y: y\in \I\} \subseteq
    \{z+j:(z,j)\in Z \times J\}.$$
Since the $t_y$'s takes at most $a/b$ different values and $x$ and $\gamma$ are
fixed, $|Z|\le (a/b)^2$. On the other hand $J$ is a proper subspace of
$\K_{\F_{p^a}}$. Thus $p^m \le |Z\times J| \le \frac{a^2 p^{m-a}}{b^2}$ and so $p^a
\le (a/b)^2 \le a^2$. This implies that $p=2$, $b=1$ and $a$ is either 2 or 4
(recall that $p$ divides $a$). Then, the index of $J$ in $\I$, as an additive
subgroup, is $|Z|=a^2=2^a$ and the elements of $Z$ form a set of representatives of
$\K$ modulo $J$. This implies that $T\circ \lambda'$ is surjective and hence we may
choose $x$ so that $\tau'_x$ is the Frobenius automorphism that maps $s\in
\F_{2^a}$ to $s^2$. We also may choose $\gamma \in \F_4\setminus \F_2$. Hence
$\gamma-\tau'_x(\gamma)=1$, so that $y\in z_y+J$, by (\ref{Igualdad}). Therefore
$y-y'\in J$ if and only if $z_y=z_{y'}$ if and only if $t_y=t_{y'}$ and $t_{\gamma
y}=t_{\gamma y'}$, for every $y,y'\in \I$. Given $u_1,u_2\in \mathcal{T}$, let
$K_{u_1,u_2}=x-u_1(x)  +\tau'_x(\gamma)(u_2(x)-x)+J$. Then
$\I/J=\{K_{u_1,u_2}:u_1,u_2\in \mathcal{T}\}$ and $y\in K_{u_1,u_2}$ if and only if
$t_y=u_2$ and $t_{\gamma y}=u_1$. Thus $\lambda\inv(P)=\{y\in \I : t_y=1
\}=\cup_{u\in \mathcal{T}} K_{u,1}$ and, by (\ref{Lambda+}), $\lambda\inv(P)$ is a
subgroup of $\I$. Hence $[\I:\lambda\inv(P)]=[\lambda\inv(P):J]=a$.

We claim that if $u+v\in \lambda\inv(P)$ and $u\not\in \lambda\inv(P)$ then $\lambda(u)=\lambda(v)$. By means of
contradiction assume that $u+v\in \lambda\inv(P)$, $u\not\in \lambda\inv(P)$ and $\lambda(u)\ne \lambda(v)$. Then
$\lambda(u+v)$ is $\F_{p^a}$-linear, so that $\lambda(u)+\lambda(v)$ is also $\F_{p^a}$-linear, by (\ref{Lambda+}).
Thus for every $\zeta\in \F_{p^a}$ and $z\in \I$ we have
    $$(\lambda(u)+\lambda(v))(\zeta z) = \zeta \lambda(u)(z)+\zeta \lambda(v)(z)$$
and
    $$(\lambda(u)+\lambda(v))(\zeta z) = \tau_u(\zeta) \lambda(u)(z)+ \tau_v(\zeta) \lambda(v)(z).$$
Therefore
    \begin{equation}\label{XYZ}
    (\gamma+\tau_u(\gamma))\lambda(u)(z)=(\gamma+\tau_v(\gamma))\lambda(v)(z).
    \end{equation}
From this equality and the assumption $\lambda(u)\ne \lambda(v)$ one deduces that $\tau_u\ne \tau_v$ and the fixed
fields of $\tau_u$ and $\tau_v$ coincides. Therefore $a=4$ and $\tau_u$ and $\tau_v$ are the two generators of
$\Gal(\F_{16}/\F_2)$. So one may assume that $\tau_u(\zeta)=\zeta^2$ and $\tau_v(\zeta)=\zeta^8$, for every $\zeta\in
\F_{16}$. Now specializing (\ref{XYZ}) to $\zeta=\gamma\in \F_4\setminus \F_2$ one deduces that $\lambda(u)=\lambda(v)$
because $\gamma+\tau_u(\gamma)=\gamma+\tau_1(\gamma)=1$. This proves the claim.

By the previous paragraph $|P|(a-1)=|Q\setminus P|=[\I:\lambda\inv(P)]-1=a-1$ and hence $P=1$. Therefore $J=0$ and
$m=a$, because $J$ has codimension $1$ as an $\F_{p^a}$-subspace of $\K$. If $a=2$ then $C$ is trivial as
affine-invariant code, by Example~\ref{Ex4}. Here we use that $b=1$, hence the defining set of $C$ is a union of
$2$-cyclotomic classes modulo $3$. Thus $a=4$ and so $\G_{a,b}=\F_{16}^*\rtimes \Gal\left(\F_{16}/ \F_2\right)$. Let
$\sigma \in\Gal(\F_{16}/\F_2)$ be the Frobenius automorphism.  Since $T\circ \lambda$ is surjective there exist
$u,v\in\I$ such that $\lambda(u)=(\gamma_1,\sigma)$ and $\lambda(v)=(\gamma_2,\sigma^2)$, for some $\gamma_i\in
\F_{16}^*$, $i=1,2$. Since $P=1$ we have that $\lambda(u+v)=(\gamma_3,\sigma^3)$ for some $\gamma_3\in\F_{16}$. Using
(\ref{Lambda+}) we conclude that every element of $\F_{16}$ is a root of the polynomial $p(X)=X+\gamma_1 X^2+ \gamma_2
X^4+\gamma_3 X^8$, which yields the desired contradiction.
\end{proof}

\begin{corollary}\label{SoloTrivial}
Let $C$ be a non-trivial affine-invariant code of length $p^m$ and $G$ be a finite group.
\begin{itemize}
\item[(a)] If $a(C)=m$, $a/b$ is coprime with $p$ and $C$ is a left $G$-code then $G$ is isomorphic to the
$p$-elementary abelian group of order $p^m$.
\item[(b)] If $a(C)=m$ and $C$ is a $G$-code then $G$ is isomorphic to the $p$-elementary abelian group of order $p^m$.
\end{itemize}
\end{corollary}

\begin{proof}
If $a(C)=m$ then $\GL(\K_{\F_{p^a}})=\GL(\K_{\K})\simeq \K^*$, a group of order coprime with $p$. Therefore, if
$\alpha:\I\rightarrow \GL(\K_{\K})$ satisfies condition (\ref{IYB}) then $\alpha(x)=1$ for every $x\in \I$. Now (b)
follows from statement (b) of Theorem~\ref{affinegroupcodes}. The proof of (a) is similar using statement (a) of
Theorem~\ref{affinegroupcodes} and the fact that the order of $\G_{a,b}$ is $|\GL(\K_{\F_{p^a}})|a/b$, which is coprime
with $p$ under the assumptions of (a).
\end{proof}

\begin{remark}{\rm
For every $\alpha:\I\rightarrow \G_{a,b}$ satisfying (\ref{IYB}), consider $\I_{\alpha}$ acting on $\I$ by conjugation
inside $\I\rtimes \G_{a,b}$. Then the map $\pi:(x,\alpha(x)\inv)\mapsto x$ is a bijective 1-cocycle, i.e.
$\pi(gh)=\pi(g)+g(\pi(h))$, for every $g,h\in \I_{\alpha}$. Groups acting on abelian groups with bijective 1-cocycles
have received the attention of several authors by its connections with the set-theoretical solutions of the Yang-Baxter
equation \cite{CJR,ESS}. }
\end{remark}

\section{A class of group code structures}\label{Ejemplos}

Theorem~\ref{affinegroupcodes} describes all the (left) group code structures of a
non-trivial affine-invariant code. The most obvious one is obtained for $\alpha$
the trivial map $x\mapsto 1$. In this case $\I_{\alpha}\simeq \I$ and this yields
the $p$-elementary abelian group structure mentioned in the introduction. In this
section we exhibit a family of other group code structures on a fixed
affine-invariant code $C$ of length $p^m$ under the assumption that $a(C)\ne m$.

We keep the notation of the previous sections, that is $\F$ and $\K=\F_{p^m}$ are finite fields of characteristic $p$
and $\I$ is the underlying additive group of $\K$. Let $C$ be a non-trivial affine-invariant code inside $\F\I$ and set
$a=a(C)$ and $b=b(C)$. Let $f$ and $\chi$ be as follows
\begin{equation}\label{Cond}
\matriz{{ll}
\cdot & f:\K \rightarrow \K \text{ is an } \F_{p^a}\text{-linear map}, \\
\cdot &  \chi:\K \rightarrow \F_{p^a} \text{ is an additive map (i.e. $\F_p$-linear)}, \\
\cdot &  \chi\ne 0\ne f, \quad f^2=0 \quad \text{ and }\quad \chi\circ f=0.}
\end{equation}
Observe that there are $\chi$ and $f$ satisfying (\ref{Cond}) if and only if $a<m$.

Consider the map
    $$\matriz{{rcl}
    \alpha=\alpha_{\chi,f}:\I & \longrightarrow & \G_{a,b} \\  x & \mapsto & 1+\chi(x)\cdot f}$$
For $x,y\in \I$
  $$\matriz{{rcl}
  \alpha\left(\alpha(y)(x)\right) &=&
  1+\left[\chi(x)+\chi\left(\chi(y)\cdot f(x)\right)\right]\cdot f\\
  &=& 1+\chi(x)\cdot f=\alpha(x)}$$
  and hence
  $$\begin{array}{l}
  \alpha\left(\alpha(y)(x)\right)\alpha(y)=\\
  \alpha(x)\alpha(y)=\left(1+\chi(x)\cdot f\right)\left(1+\chi(y)\cdot f\right)=\\
  1+\chi(y)\cdot f+\chi(x)\cdot f=\alpha(x+y).
  \end{array}$$
Therefore $\alpha$ satisfies condition (\ref{IYB}). Since $\alpha(x)\inv=1-\chi(x)f$, by
Theorem~\ref{affinegroupcodes}, $C$ is a left $\I_{\chi,f}$-code, where
    $$\I_{\chi,f}=\{(x,1-\chi(x)f):x\in \I\}.$$
Moreover, using the notation of Theorem~\ref{affinegroupcodes}, we have $\beta(x,y)=\alpha(x)^{-1}(y)-y=-\chi(x)f(y)$
and hence, if $\chi$ is $\F_{p^a}$-linear then $C$ is $\I_{\chi,f}$-code.

Our next goal consist in describing the structure of $\I_{\chi,f}$. For that we need to introduce some group
constructions from some vector spaces.

\begin{notation}\label{FF}
Given an $\F_p$-vector space $V$ and linear map $\mu:V\rightarrow \F_{p^a}$, we define the group $V_{\mu}=V\times V$
with the following product:
    $$(v_1,w_1)(v_2,w_2)=(v_1+v_2-\mu(w_1)w_2,w_1+w_2),$$
for $v_1,w_1,v_2,w_2\in V$.

If $U$ is an additive subgroup of $\F_{p^a}$ then we consider the following action on $V_{\mu}$:
    $$u(v,w)=(v-uw,w), \quad (u\in U, v,w\in V).$$
The corresponding semidirect product is denoted by $V_{\mu}\rtimes U$ and its elements by $(v_1,v_2;u)$, with
$v_1,v_2\in V$ and $u\in U$.
\end{notation}

\begin{lemma}\label{abelian}
{\rm If $(x,f)$ and $(y,g)$ belong to $\I\rtimes \G_{a,b}$ then
  $$(x,f)(y,g)=(y,g)(x,f) \Leftrightarrow \left\{\matriz{{l} x+f(y)= y+g(x) \\\text{ and }
  fg=gf.}\right.$$}
\end{lemma}

\begin{proof}
Straightforward.
\end{proof}

\begin{theorem}\label{lf}
Let $\chi$ and $f$ be as in (\ref{Cond}) and consider $\K$ as an $\F_p$-vector space. Then
\begin{enumerate}[(a)]
\item\label{Abelian}
$\I_{\chi,f}$ is abelian if and only if $\chi$ is $\F_{p^a}$-linear and $\ker(\chi)\subseteq \ker(f)$. In this case
$\ker(\chi)=\ker(f)$.
\item
If $\I_{\chi,f}$ is non-abelian then the center of $\I_{\chi,f}$ is $\{(z,1): z\in \ker(f)\cap \ker(\chi)\}$.
\item\label{Odd}
If $p$ is odd then $\I_{\chi,f}$ has exponent $p$. If $p=2$ then the exponent of $\I_{\chi,f}$ is $4$.
\item\label{Two} Let $V=\Imagen(f)$, $Z$ a
complement of $V$ in $\ker(\chi)\cap \ker(f)$ and $U$ a complement of $\ker(\chi)\cap \ker(f)$ in $\ker(f)$. Then
    $$\I_{\chi,f} \simeq Z \times (V_{\chi \circ g}\rtimes \chi(U)).$$
where $g:V\rightarrow \K$ is an additive map satisfying $f\circ g = 1_V$ and $g f(\ker(\chi))\subseteq \ker(\chi)$.
\end{enumerate}
\end{theorem}

\begin{proof}
(a) Using Lemma~\ref{abelian}, it is easy to see that $(x,\alpha(x)\inv)$ and $(y,\alpha(y)\inv)$ commute if and only
if $\chi(x)f(y)=\chi(y)f(x)$. Using this and the assumption $\chi\ne 0$, one easily follows that if $\I_{\chi,f}$ is
abelian then $\ker(\chi)\subseteq \ker(f)$. Furthermore, if $\gamma\in \F_{p^a}$ and $x,y\in \K$ then $\chi(\gamma
y)f(x) = \chi(x)f(\gamma y) = \gamma \chi(x)f(y) = \gamma \chi(y)f(x)$. Using that $f\ne 0$ one deduces that $\chi$ is
$\F_{p^a}$-linear. Therefore, $\ker(\chi)=\ker(f)$ because $\ker(\chi)$ has codimension $1$ in $\K_{\F_{p^a}}$ and
$\ker(f)$ is a proper subspace of $\K_{\F_{p^a}}$.

Conversely, assume that $\chi$ is $\F_{p^a}$-linear and $\ker(\chi)\subseteq \ker(f)$. Then the equality holds as
above. Let $v\in \K\setminus \ker(\chi)$ and for every $x,y\in\K$ write $x=w_x+\beta_x v$ and $y=w_y+\beta_y v$ with
$w_x,w_y\in \ker(\chi)$ and $\beta_x,\beta_y\in \F_{p^a}$. Then $\chi(x)f(y)=\beta_x \beta_y \chi(v)f(v) = \chi(y)f(x)$
and hence $\left(x,\alpha(x)^{-1}\right)$ and $\left(y,\alpha(y)^{-1}\right)$ commute.

(b) Assume that $\I_{\chi,f}$ is non-abelian. By Lemma~\ref{abelian}, $(z,\alpha(z)\inv)$ belongs to the center of
$\I_{\chi,f}$ if and only if $\chi(x)f(z)=\chi(z)f(x)$, for every $x\in \K$. In particular, if $z\in \ker(f)\cap
\ker(\chi)$ then $(z,\alpha(z)\inv)=(z,1)$ belongs to the center of $\I_{\chi,f}$. Conversely, let
$v=(z,\alpha(z)\inv)$ belong to the center of $\I_{\chi,f}$. If $x_1\in \K\setminus \ker(\chi)$ and $x_2\in \K\setminus
\ker(f)$ then from the equalities $\chi(x_1)f(z)=\chi(z)f(x_1)$ and $\chi(x_2)f(z)=\chi(z)f(x_2)$ one deduces that
$z\in \ker(\chi)$ if and only if $z\in \ker(f)$. So it is enough to show that $f(z)=0$ or $\chi(z)=0$. By (a), either
$\chi$ is not $\F_{p^a}$-linear or there is $y\in \ker(\chi)\setminus \ker(f)$. In the latter case,
$0=\chi(y)f(z)=\chi(z)f(y)$ and hence $\chi(z)=0$. In the former case, there is $\gamma\in \F_{p^a}$ and $x\in \K$ with
$\chi(\gamma x)\ne \gamma \chi(x)$. However, $\chi(\gamma x)f(z) = \chi(z)f(\gamma x) = \gamma \chi(z)f(x) = \gamma
\chi(x)f(z)$ and hence $f(z)=0$.

(c) Let $x\in \K$ and $v=(x,\alpha(x)\inv)=(x,1-\chi(x)f)\in \I_{\chi,f}$. A straightforward calculation shows that
    $$v^p = \left(-\sum_{i=0}^{p-1} i\chi(x)f(x),1\right)
    = \left(\frac{p(p-1)}{2} \chi(x)f(x),1\right).$$
Hence, if $p$ is odd then $v^p=(0,1)$ and if $p=2$ then $v^2\in \I$, so that $v^4=1$. By means of contradiction, assume
that $p=2$ and the exponent of $\I_{\chi,f}$ is $2$. Then $\chi(x)f(x)=0$, for every $x\in \K$. In particular, $f(x)=0$
for every $x\in \K\setminus \ker(\chi)$. If $w\in \ker(\chi)$ and $x\in \K\setminus \ker(\chi)$ then
$f(w)=f(x+w)-f(x)=0$. This shows that $f=0$, a contradiction.

(d) The existence of the map $g$ follows by standard linear algebra arguments.

Let $W=g(f(\ker(\chi)))$ and $W'=g(H)$, where $H$ is a complement of $f(\ker(\chi))$ in $V$. Clearly the restriction
maps $f:W\oplus W' \rightarrow V$ and $g:V\rightarrow W\oplus W'$ are mutually inverse to each other.

Since $f\circ g = 1_V$, we have $W\cap \ker(f)=0$. Furthermore, $\dim(\ker(\chi)) = \dim(\ker(\chi)\cap \ker(f))+\dim
f(\ker(\chi)) = \dim(\ker(\chi)\cap \ker(f))+\dim(W)$. This shows that $\ker(\chi)=W\oplus (\ker(\chi)\cap \ker(f))$.

We claim that $\K=W'\oplus (\ker(\chi)+\ker(f))$. Indeed, if $x\in W'\cap \left(\ker(\chi) + \ker(f)\right)$ then
$x=u+v=g(h)$ for some $u\in\ker(\chi)$, $v\in\ker(f)$ and $h\in H$. So, $f(x)=f(u)=fg(h)=h$, and then $h\in
f\left(\ker(\chi)\right)\cap H=\{0\}$. Thus $x=0$. Moreover,
    $$\matriz{{rcl}
    \dim \K &=& \dim(V)+\dim(\ker(f)) \\
    &=&\dim(V)+\dim(\ker(f)+\ker(\chi))+ \\ && \dim(\ker(\chi)\cap \ker(f)) - \dim(\ker(\chi)) \\
    &=& \dim(V)- \dim(f(\ker(\chi))) + \\ && \dim(\ker(f)+\ker(\chi)) \\
    &=& \dim(H) + \dim(\ker(f)+\ker(\chi)) \\
    &=& \dim(W') + \dim(\ker(f)+\ker(\chi)).}$$
This proves the claim.

Then $\K=Z\oplus V \oplus W \oplus W' \oplus U$ and the product in $\I_{\chi,f}$ is given by
    $$\matriz{{ll}(x_1,1-\chi(x_1)f)(x_2,1-\chi(x_2)f)
    = \\(x_1+x_2-\chi(w_1'+u_1)f(w_2+w_2'),
    1-\chi(x_1+x_2)f)}$$
for $x_1=z_1+v_1+w_1+w_1'+u_1$ and $x_2=z_2+v_2+w_2+w_2'+u_2$, with $z_i\in Z, v_i\in V, w_i\in W, w'_i\in W'$ and
$u_i\in U$. We conclude that the map $\I_{\chi,f} \rightarrow Z \times (V_{\chi\circ g}\rtimes \chi(U))$ given by
    $$\matriz{{r}
    (z+v+w+w'+u,1-\chi(w'+u)f)\mapsto \\
    \left(z,\left(v,f(w+w');\chi(u)\right)\right)}$$
is a bijection. The fact that this map is a group homomorphism follows by straightforward
computations.
\end{proof}
\bigskip

As it was mentioned above, if $\chi$ is $\F_{p^a}$-linear then $C$ is $\I_{\chi,f}$-code. In this case one can obtain a
more friendly description of $\I_{\chi,f}$.

Let $\mathcal{F}_a=(\F_{p^a})_1$, i.e. $\mathcal{F}_a = \F_{p^a}\times \F_{p^a}$ with the product
$(x_1,y_1)(x_2,y_2)=(x_1+y_1+x_1y_2,x_2+y_2)$ (see Notation~\ref{FF})l. For an $\F_{p^a}$-vector space $V$ consider the
following action of $\mathcal{F}_a$ on $V\times V$:
    \begin{equation}\label{Action}
    (x,y)\cdot (u,v) = (u-yv,v), \quad (x,y\in \F_{p^a}, u,v \in V).
    \end{equation}
Let $(V\times V)\rtimes \mathcal{F}_a$ denote the corresponding semidirect product and
denote its elements with $(v_1,v_2;x_1,x_2)$ for $v_i\in V$ and $x_i\in \F_{p^a}$.

\begin{corollary}\label{Estr}
Let $\chi$ and $f$ be as in (\ref{Cond}) and assume that $\chi$ is $\F_{p^a}$-linear. Let $u$ denote the rank of $f$.
Then we have:
\begin{enumerate}[(a)]
\item If $\ker(\chi)=\ker(f)$ and $p$ is odd then $\I_{\chi,f}$ is $p$-elementary abelian.
\item If $\ker(\chi)=\ker(f)$ and $p=2$ then $\I_{\chi,f}$ is a direct product of $m-2a$ copies of groups of order $2$ and $a$ copies of cyclic groups of order $4$.
\item If $\ker(f)\not\subseteq \ker(\chi)$ then $\I_{\chi,f}$ is isomorphic to the group
    $$\F_{p^a}^{\frac{m}{a}-2u-1}\times \left(\left(\F_{p^a}^u \times \F_{p^a}^u\right)\rtimes \F_{p^a}\right).$$
\item If $\ker(f)\subsetneq \ker(\chi)$ then $\I_{\chi,f}$ is isomorphic to the group
    $$\F_{p^a}^{\frac{m}{a}-2u}\times \left(\left(\F_{p^a}^{u-1} \times \F_{p^a}^{u-1})\right)\rtimes \mathcal{F}_a\right).$$
\end{enumerate}
\end{corollary}

\begin{proof}
(a) and (b). Assume that $\ker(\chi)=\ker(f)$. Then $\I_{\chi,f}$ is abelian, by statement (\ref{Abelian}) of
Theorem~\ref{lf}. If $p$ is odd then $\I_{\chi,f}$ is elementary abelian by statement~(\ref{Odd}) of the same
proposition. Suppose that $p=2$. Then $\I_{\chi,f}$ is a direct product of cyclic groups of order $2$ or $4$, by
statement~(\ref{Two}) of Theorem~\ref{lf}. If $x\in \I$ then $v=(x,1-\chi(x)f)$ has order $\le 2$ if and only if
$(0,1)=v^2=(-\chi(x)f(x),1)$ if and only if $x\in \ker(\chi)$. Thus, if $\I_{\chi,f}$ is a direct product of $k$ copies
of groups of order $2$ and $l$ copies of cyclic groups of order $4$ then $k+l=a\dim_{\F_{2^a}}(\ker(\chi))=m-a$ and
$m=k+2l$. Solving this two equations we deduce that $k=m-2a$ and $l=a$.

In the remainder of the proof we use the notation of Theorem~\ref{lf} and its proof. So $\I_{\chi,f} \simeq Z \times
(V_{\chi \circ g}\rtimes \chi(U))$. Notice that $Z,V,U,W$ and $W'$ can be selected as $\F_{p^a}$-subspaces of $\K$.
Since $\ker(\chi)$ has codimension 1 in $\K_{\F_{p^a}}$ and $U\oplus W'$ is a complement of $\ker(\chi)$ in $\K$,
either $U=0$ or $W'=0$.

(c) Suppose that $\ker(f)\not\subseteq \ker(\chi)$. Then $W'=0$, $\chi(U)=\F_{p^a}$ and $\chi\circ g=0$. By statement
(d) of Theorem~\ref{lf}, $\I_{\chi,f}\simeq Z \times ((V\times V) \rtimes \F_{p^a})$. Furthermore $V\simeq \F_{p^a}^u$
and $Z\simeq \F_{p^a}^{\frac{m}{a}-2u-1}$.

(d) Assume now that $\ker(f)\subsetneq \ker(\chi)$. Then $U=0$ and $W$ and $W'$ are $\F_{p^a}$-subspaces of $\K$ of
dimensions $u-1$ and $1$ respectively. Using this one deduces that $X=\{(v_1,v_2):v_1,v_2\in f(W)\}$ and
$X'=\{(v'_1,v'_2):v'_1,v'_2\in f(W')\}$ are subgroups of $V_{\chi \circ g}$ with $X\cap X'=1$ and $X$ normal in
$V_{\chi \circ g}$. Then $V_{\chi \circ g}=X\rtimes X'$, so that $\I_{\chi,f}\simeq Z \times (X\rtimes X')$. Moreover
$X= f(W)\times f(W)$, because $\chi\circ g$ vanishes on $f(W)$. On the other hand, the map $\phi:X'\rightarrow
\mathcal{F}_a$, given by $\phi((v_1,v_2))=(\chi g(v_1),\chi g(v_2))$, is a group isomorphism. Hence the action of $X'$
on $X$ by conjugation yields and action of $\mathcal{F}_a$ on $X$ via $\phi$. It is easy to see that this action is
precisely the action defined in (\ref{Action}). So $\I_{\chi,f}\simeq Z\times ((f(W)\times f(W))\rtimes
\mathcal{F}_a)$. Finally, $Z\simeq \F_{p^a}^{\frac{m}{a}-2u}$ and $f(W)\simeq \F_{p^a}^{u-1}$.
\end{proof}
\bigskip

In the remainder of the section we fix a non-trivial affine-invariant code $C$ with $a=a(C)\ne m$ and show how to
obtain $\F_{p^a}$-linear maps $\chi$ and $f$ satisfying (\ref{Cond}) and yielding all the cases of
Corollary~\ref{Estr}. For that we start with an arbitrary non-zero linear form $\chi$ of $\K_{\F_{p^a}}$ and construct
an endomorphism $f$ of $\K_{\F_{p^a}}$ satisfying the conditions of (\ref{Cond}). The existence of the endomorphism $f$
in all the cases is clear.

To obtain an abelian group code structure on $C$ with a given non-zero linear form $\chi$ we just need an endomorphism
$f$ of $\K_{\F_{p^a}}$ with $\ker(f)=\ker(\chi)$ and $0\ne f(v) \in \ker(\chi)$ for a given $v\in \K\setminus
\ker(\chi)$.

If $m>2a$ then it is always possible to obtain a non-abelian group code structure on $C$. In fact, for every positive
integer $u$ with $2u\le \frac{m}{a}-1$ there are endomorphisms $f_1$ and $f_2$ of $\K_{\F_{p^a}}$, satisfying the
conditions of (\ref{Cond}) such that $\I_{\chi,f_1}$ and $\I_{\chi,f_2}$ are as in statements (c) and (d) of
Corollary~\ref{Estr} respectively. Indeed, in this case, the dimension of $\ker(\chi)$ as $\F_{p^a}$-vector space is
$\frac{m}{a}-1\ge 2u$. Thus we have $\K=\ker(\chi)\oplus X$ and $\ker(\chi)=Z_1\oplus V \oplus W_1=Z_2\oplus V \oplus
W_2$, for $\F_{p^a}$-subspaces $X,Z_1,Z_2,V,W_1$ and $W_2$ of $\K$, where $\dim_{\F_{p^a}}(X)=1$,
$\dim_{\F_{p^a}}(V)=\dim_{\F_{p^a}}(W_1)=u$ and $\dim_{\F_{p^a}}(W_2)=u-1$. Then we can construct the desired
endomorphisms $f_1$ and $f_2$ of $\K$ by setting $f_i(W_i\oplus X)=V$ for $i=1,2$, $\ker(f_1)=Z_1\oplus V \oplus X$ and
$\ker(f_2)=Z_2\oplus V$. Observe that, in case (d), we have $u>1$. On the other hand, if $u=1$ then
$\ker(f_2)=\ker(\chi)$ and hence $\I_{\chi,f_2}$ is abelian.

However if $m=2a$ then it is not possible to obtain a non-abelian group code structure by the following result.

\begin{corollary}
Let $C$ be a non-trivial affine-invariant code of length $p^m$. Then the following conditions are equivalent.
\begin{itemize}
\item[(a)] $C$ is a $G$-code for some non-abelian group $G$.
\item[(b)] $2a(C)<m$.
\end{itemize}
Furthermore, if $2a(C)\ge m$ and $C$ is a $G$-code then either $G\simeq \I$ or $p=2$ and $G$ is a direct product of $a$
copies of cyclic groups of order $4$.
\end{corollary}

\begin{proof}
(b) implies (a) is a consequence of the arguments given before the corollary.

(a) implies (b) Let $a=a(C)$ and assume that $2a\ge m$, so that $m$ is either $a$ or $2a$ and $G$ is a group such that
$C$ is a $G$-code. By Theorem~\ref{affinegroupcodes}, $G$ is isomorphic to $\I_{\alpha}$, for $\alpha:\I\rightarrow
\GL(\K_{\F_{p^a}})$ a map satisfying condition (\ref{IYB}) such that $\beta(x,y)=\alpha(x)\inv(y)-y$ is
$\F_{p^a}$-bilinear. If $\alpha(x)=1$ for every $x$ then $G\simeq \I$ and so $G$ is abelian as wanted. This happens,
for example, if $m=a$ because in this case the order of $\GL(\K_{\F_{p^a}})=K^*$ is coprime with $p$ and the order of
$\alpha(x)$ is a $p$-th power. Assume now that $\alpha(x)\ne 1$ for some $x\in \I$. Then $m=2a$ and, by Sylow's
Theorem, we may assume that $\alpha(x)$ belongs to a prescribed Sylow $p$-subgroup of $\GL(\K_{\F_{p^a}})$. For
example, we may fix a basis $u_1,u_2$ of $\K_{\F_{p^a}}$ and assume that $\alpha(x)(u_1)=u_1$ and $\alpha(x)(u_2)-u_2 =
\chi(x)u_1$ for some $\chi(x)\in \F_{p^a}$ (see the proof of Lemma~\ref{suma imagenes}). Let $f$ be the
$\F_{p^a}$-linear endomorphism of $\K$ given by $f(u_1)=0$ and $f(u_2)=u_1$. Then $\alpha(x)\inv=1-\chi(x)f$ and
$\beta(x,y)=-\chi(x)f(y)$. Since $\beta$ is $\F_{p^a}$-linear, $\chi$ is $\F_{p^a}$-bilinear. Furthermore,
$$\matriz{{l}
1+(\chi(x)+\chi(y))f=1+\chi(x+y)f=\alpha(x+y)=\\
\alpha(\alpha(y)(x))\alpha(y)=(1+\chi(x+\chi(y)f(x))f)(1+\chi(y)f)=\\
1+(\chi(x)+\chi(y))f+\chi(y)\chi(f(x))f}
$$
and we conclude that $\chi\circ f=0$, i.e. $\chi$ and $f$ satisfy the conditions of (\ref{Cond}) and $G\simeq
\I_{\chi,f}$. Moreover $\dim_{\F_{p^a}}(\ker(\chi))=1$ and so $\ker(f)=f(\K)=\F_{p^a}v_1=\ker(\chi)$. We conclude that
$\I_{\chi,f}$ is abelian, by statement~(\ref{Abelian}) of Theorem~\ref{lf}.

Finally, the last statement is a consequence of statements (a) and (b) of Corollary~\ref{Estr}.
\end{proof}

\end{document}